\documentclass[onecolumn]{IEEEtran}

\usepackage{graphicx}
\usepackage{color}
\usepackage{amssymb}
\usepackage{dashbox}
\usepackage{amsmath}
\usepackage{epstopdf}
\usepackage[noadjust]{cite}
\usepackage{comment}
\usepackage{multirow}

\newcommand{\DHF}{\mbox{$\mathsf{DHF}$}}
\newcommand{\PHF}{\mbox{$\mathsf{PHF}$}}
\newcommand{\HF}{\mbox{$\mathsf{HF}$}}
\newcommand{\rHF}{\mbox{$\mathsf{HF}^\circ$}}
\newcommand{\rSHF}{\mbox{$\mathsf{SHF}^\circ$}}
\newcommand{\SHF}{\mbox{$\mathsf{SHF}$}}

\newtheorem{theorem}{Theorem}

\newtheorem{lemma}{Lemma}

\begin{document}

\title{Hierarchical Recovery in Compressive Sensing}

\author{
Charles J.\ Colbourn, Daniel Horsley, and Violet R.\ Syrotiuk,~\IEEEmembership{Senior Member,~IEEE}
%
%
\thanks{C.~J.\ Colbourn and V.~R.\ Syrotiuk are with the School of Computing, Informatics, and Decision Systems Engineering, Arizona State University, Tempe, AZ, U.S.A., 85287-8809, {\tt \{colbourn,syrotiuk\}@asu.edu}}
\thanks{D. Horsley is with the School of Mathematical Sciences, Monash University, Vic 3800, Australia, {\tt daniel.horsley@monash.edu}}
}


\maketitle


\begin{abstract} 
A combinatorial approach to compressive sensing based on a deterministic column replacement technique is proposed.
Informally, it takes as input a pattern matrix and ingredient measurement matrices, and results in a larger measurement matrix by replacing elements of the pattern matrix with columns from the ingredient matrices.
This hierarchical technique yields great flexibility in sparse signal recovery.
Specifically, recovery for the resulting measurement matrix does not depend on any fixed algorithm but rather on the recovery scheme of each ingredient matrix.
In this paper, we investigate certain trade-offs for signal recovery, considering the computational investment required.
Coping with noise in signal recovery requires additional conditions, both on the pattern matrix and on the ingredient measurement matrices.
\end{abstract}



\section{Introduction}\label{sec:intro}

Nyquist's sampling theorem provides a sufficient condition for full recovery of a band-limited signal: sample the signal at a rate that is twice the band-limit.
However, there are cases when full recovery may be achieved with a sub-Nyquist sampling rate.
This occurs with signals that are sparse (or compressible) in some domain, such as those that arise in applications in sensing, imaging, and communications, and has given rise to the field of \emph{compressive sensing} \cite{candes06-1,Baraniuk}  (also called \emph{compressive sampling}).

Consider the following framework for compressive sensing.
An \emph{admissible signal} of \emph{dimension} $n$ is a vector in ${\mathbb R}^n$ that is known {\sl a priori} to be taken from a given set $\Phi \subseteq {\mathbb R}^n$.
A \emph{measurement matrix} $A$ is a matrix from ${\mathbb R}^{m \times n}$.
\emph{Sampling} a signal $\mathbf{x} \in {\mathbb R}^n$ corresponds to computing the product $A \mathbf{x} = \mathbf{b}$.
Once sampled, \emph{recovery} involves determining the unique signal $\mathbf{x} \in \Phi$ that satisfies $A \mathbf{x}=\mathbf{b}$ using only $A$ and $\mathbf{b}$.
If $\Phi = {\mathbb R}^n$, recovery can be accomplished only if $A$ has rank $n$, and hence $m \geq n$.
However for more restrictive admissible sets $\Phi$, recovery may be accomplished when $m < n$.

Given a measurement matrix $A$, an equivalence relation $\equiv_A$ is defined so that for signals $\mathbf{x}, \mathbf{y} \in  {\mathbb R}^n$, we have $\mathbf{x} \equiv_A \mathbf{y}$ if and only if $A \mathbf{x} = A \mathbf{y}$.
If for every equivalence class $P$ under $\equiv_A$, the set $P \cap \Phi$ contains at most one signal then in principle recovery is possible.
Because $A \mathbf{x} = A \mathbf{y}$ ensures that $A(\mathbf{x}-\mathbf{y}) = \mathbf{0}$, this can be stated more simply:
An equivalence class $P$ of $\equiv_A$ can be represented as $\{ \mathbf{x} + \mathbf{y} : \mathbf{y} \in N(A)\}$ for any $\mathbf{x} \in P$, where $N(A)$ is the \emph{null space} of $A$, i.e., the set $\{ \mathbf{x} \in {\mathbb R}^n: A \mathbf{x} = \mathbf{0}\}$.
Recoverability is therefore equivalent to requiring that, for every signal $\mathbf{x} \in \Phi$, there is no $\mathbf{y} \in N(A) \setminus \{\mathbf{0}\}$ with $\mathbf{x} + \mathbf{y} \in \Phi$.

In order to make use of these observations, a reasonable {\sl a priori} restriction on the signals to be sampled is identified, suitable measurement matrices with $m \ll n$ are formed, and a reasonably efficient computational strategy for recovering the signal is provided.
A signal is $t$-\emph{sparse} if at most $t$ of its $n$ coordinates are nonzero.
The recovery of  $t$-sparse signals is the domain of \emph{compressive sensing}.
An admissible set of signals $\Phi$ has \emph{sparsity} $t$ when every signal in $\Phi$ is $t$-sparse.
An admissible set of signals $\Phi$ is $t$-\emph{sparsifiable} if there is a full rank matrix $B \in {\mathbb R}^{n \times n}$ for which $\{ B \mathbf{x} : \mathbf{x} \in \Phi\}$ has sparsity $t$.
We assume throughout that when the signals are sparsifiable, a change of basis $B$ is applied so that the admissible signals have sparsity $t$.

A measurement matrix \emph{has $(\ell_0,t)$-recoverability} when it permits exact recovery of all $t$-sparse signals.
A basic problem is to design measurement matrices with $(\ell_0,t)$-recoverability where $m \ll n$ such  that recovery can be accomplished efficiently.
Suppose that measurement matrix $A$ has $(\ell_0,t)$-recoverability.
Then in principle, given $A$ and $\mathbf{b}$, recovery of the signal $\mathbf{x}$ can be accomplished by solving the $\ell_0$-minimization problem $\min\{ ||\mathbf{x}||_0 : A \mathbf{x} = \mathbf{b}\}$.
To do so the possible supports of signals from fewest nonzero entries to most are first listed.
For each, reduce $A$ to $A'$ and $\mathbf{x}$  to $\mathbf{x}'$ by eliminating coordinates in the signal assumed to be zero.
Examine the now overdetermined system $A' \mathbf{x}' = \mathbf{b}$.
When equality holds, a solution is found;  we are guaranteed to find one by considering all possible supports with at most $t$ nonzero entries.
Such an enumerative strategy is prohibitively time-consuming, examining as many as $\binom{n}{t}$ linear systems when the signal has sparsity $t$.
Natarajan \cite{Natarajan} showed that we cannot expect to find a substantially more efficient solution, because the problem is NP-hard.

Instead of the $\ell_0$-minimization problem, Chen, Donoho, Huo, and Saunders \cite{ChenDS,Donoho01} suggest considering the $\ell_1$-minimization problem $\min\{ ||\mathbf{x}||_1 : A \mathbf{x} = \mathbf{b} \}$.
While this can be solved using standard linear programming techniques, to be effective it is necessary that for each $t$-sparse signal $\mathbf{x}$, the unique solution to $\min\{ ||\mathbf{z}||_1 : A \mathbf{z} = A \mathbf{x} \}$ is $\mathbf{x}$.
This property is  $(\ell_1,t)$-recoverability.
A  necessary and sufficient condition for $(\ell_1,t)$-recoverability has been explored, beginning with Donoho and Huo \cite{Donoho01} and subsequently in \cite{Stojnic,Zhang08,EladBruckstein,Fuchs2004,Fuchs2005,Tropp2005,GribonvalMorten2003}.

A measurement matrix $A$ meets the $(\ell_0,t)$-\emph{null space condition}  if and only if $N(A)\setminus \{\mathbf{0}\}$ contains no $(2t)$-sparse vector.
For $\mathbf{y} \in {\mathbb R}^n$ and $C \subset \{1,\dots,n\}$, define $\mathbf{y}_{|C} \in {\mathbb R}^n$ to be the vector such that $(y_{|C})_\gamma = y_{\gamma}$ if $\gamma \in C$ and $(y_{|C})_\gamma=0$ otherwise.
A measurement matrix $A$ meets the $(\ell_1,t)$-\emph{null space condition} if and only if for every $\mathbf{y} \in N(A) \setminus \{\mathbf{0}\}$ and every $C \subset \{1,\dots,n\}$ with $|C|=t$, $||\mathbf{y}_{|C}||_1 < \frac{1}{2} ||\mathbf{y}||_1$. \\

\begin{lemma}\label{lem:nsl0} {\rm (\cite{CHM-TCom}, for example)}
Measurement matrix $A \in {\mathbb R}^{m \times n}$ has $(\ell_0,t)$-recoverability if and only if $A$ meets the $(\ell_0,t)$-null space condition.
\end{lemma}

\hspace*{0.5in} \\

\begin{lemma}\label{lem:nsl1} {\rm (\cite{Zhang08}, for example)}
Measurement matrix $A \in {\mathbb R}^{m \times n}$ has $(\ell_1,t)$-recoverability if and only if $A$ meets the $(\ell_1,t)$-null space condition.
\end{lemma}

\hspace*{0.3in} \\

To establish $(\ell_1,t)$-recoverability, and hence also $(\ell_0,t)$-recoverability, Cand\`es and Tao \cite{candes08,candes05} introduced the \emph{Restricted Isometry Property} (RIP).
For $A \in {\mathbb R}^{m\times n}$, the $d$th \emph{RIP parameter} of $A$, $\delta_d(A)$, is the smallest $\delta$ so that, for some constant $R > 0$, $(1 -\delta)  R  (||\mathbf{x}||_2)^2 \leq (||A \mathbf{x}||_2)^2 \leq (1 +\delta)  R  (||\mathbf{x}||_2)^2$, for all $\mathbf{x}$ with $||\mathbf{x}||_0 \leq d$.
The $d$th RIP parameter is better when $\delta_d(A)$ is smaller as the bounds are tighter.
The RIP parameters have been employed extensively to establish $(\ell_1,t)$-recoverability, particularly for randomly generated measurement matrices \cite{candes06-3,candes05,candes06-2}, but also for those generated using deterministic constructions \cite{Cohen,devore07-1}.
Commonly, $\delta_{2t} < \sqrt{2}-1$ is required for $(\ell_1,t)$-recoverability; see \cite{candes08} for example.
The property of $(\ell_1,t)$-recoverability in the presence of noise has also been considered.
Conditions on the RIP parameters are sufficient but in general not necessary for recoverability.

Combinatorial approaches to compressive sensing are detailed in \cite{Berinde,Graham,Gilbert,Gilbert01,Iwenpreprint,XuHassibi2007,JafarpourEtAl2009}.
We pursue a different combinatorial approach here, using a deterministic column replacement technique based on hash families.
The use of an heterogeneous hash family provides an explicit hierarchical construction of a large measurement matrix from a library of small ingredient matrices.
Strengthening hash families provide a means to increase the level of sparsity supporte, allowing the ingredient matrices to be designed for lower sparsity than the larger measurement matrix produced.

In this paper we show that the heterogeneity extends to signal recovery: it is interesting that the ingredient measurement matrices need not all employ the same recovery algorithm.
This enables hierarchical recovery for the large measurement matrix; however, this can be computationally prohibitive.
By restricting the hash family to be linear, recovery for the large measurement matrix can be achieved in sublinear time even when computationally intensive methods are used for each ingredient matrix.
To be practical, recovery methods based on hash families must deal with noise in the signal effectively.
Suitable restrictions on the hash family and on each ingredient matrix used in the hierarchical method are shown to be sufficient to permit recovery in the presence of noise.

The rest of this paper is organized as follows.
The results on homogeneous hash families in Section \ref{sec:hf-cs} demonstrate that a recovery scheme based on $(\ell_0,t)$- or $(\ell_1,t)$-recoverability can be `lifted' from the ingredient measurement matrices to the matrix resulting from column replacement.
Section \ref{sec:gen-hf} considers a generalization of hash families to allow for ingredient matrices with other recovery algorithms, and the computational investment to recover the signal.
Signal recovery without noise is considered first, and the conditions for a sublinear time recovery algorithm described.
Section \ref{sec:noise} considers the recovery of almost-sparse signals to deal with noise in the signal.
Finally, Section \ref{sec:conclusion} draws relevant conclusions.


\section{Hash Families and Compressive Sensing} \label{sec:hf-cs}

\subsection{Column Replacement and Hash Families for Compressive Sensing} \label{sec:crhf-cs}

Let $A  \in {\mathbb R}^{r \times k}$, $A = (a_{ij})$,  be an \emph{ingredient} matrix.
Let $P \in \{1,\dots,k\}^{m \times n}$, $P= (p_{ij})$, be a \emph{pattern} matrix.
The columns of $A$ are indexed by elements of $P$.
For each row $i$ of $P$, replace element $p_{ij}$ with a copy of column $p_{ij}$ of $A$.
The result is an $rm \times n$ matrix $B$, the \emph{column replacement of $A$ into $P$}.
Fig.~\ref{fig:colreplace} gives an example of column replacement.

\begin{figure}[h]
\begin{minipage}[b]{0.3\linewidth}
\begin{displaymath}
P = \left[ \begin{array}{c}
{\tt 1231} \\
{\tt 3121} \\
\end{array} \right]
\end{displaymath}
\end{minipage}
\begin{minipage}[b]{0.3\linewidth}
\begin{displaymath}
A = \left[ \begin{array}{c}
{\tt a}_{11}{\tt a}_{12}{\tt a}_{13} \\
{\tt a}_{21}{\tt a}_{22}{\tt a}_{23} \\
\end{array} \right]
\end{displaymath}
\end{minipage}
\begin{minipage}[b]{0.3\linewidth}
\begin{displaymath}
B = \left[ \begin{array}{c}
{\tt a}_{11}{\tt a}_{12}{\tt a}_{13}{\tt a}_{11} \\
{\tt a}_{21}{\tt a}_{22}{\tt a}_{23}{\tt a}_{21} \\ 
{\tt a}_{13}{\tt a}_{11}{\tt a}_{12}{\tt a}_{11} \\
{\tt a}_{23}{\tt a}_{21}{\tt a}_{22}{\tt a}_{21} \\
\end{array} \right]
\end{displaymath}
\end{minipage}
\caption{$B$ is the column replacement of $A$ into $P$. \label{fig:colreplace}}
\end{figure}

When the ingredient matrix $A$ is a measurement matrix that meets one of the null space conditions for a given sparsity, our interest is to ensure that the sparsity supported by $B$ is at least that of $A$.
Not every pattern matrix $P$ suffices for this purpose.
Therefore, we examine the requirements on $P$.

Let $m$, $n$, and $k$ be positive integers.
An \emph{hash family} \HF$(m; n, k)$, $P = (p_{ij})$, is an $m \times n$ array, in which each cell contains one symbol from a set of $k$ symbols.
An hash family is \emph{perfect} of \emph{strength} $t$, denoted \PHF$(m; n, k, t)$, if in every $m \times t$ subarray of $P$ at least one row consists of distinct symbols; see \cite{Alon86,STW}.
Fig.~\ref{fig:PHF} gives an example of a perfect hash family \PHF$(6; 12, 3, 3)$. For example, for the $6 \times 3$ subarray involving columns $4, 5,$ and $6$, only the fourth row consists of distinct symbols.

\begin{figure} [h]
\centering
\begin{tabular}{ccccccccccccc}
& & & & $\downarrow$ & $\downarrow$ & $\downarrow$ & & & & & & \\
& 0 & 1 & 2 & 2 & 1 & 2 & 2 & 0 & 1 & 1 & 0 & 0 \\
& 0 & 2 & 1 & 0 & 2 & 2 & 2 & 1 & 0 & 1 & 2 & 1 \\
& 1 & 0 & 0 & 2 & 2 & 2 & 1 & 1 & 2 & 1 & 0 & 2 \\
$\rightarrow$ & 2 & 0 & 1 & \framebox{1} & \framebox{2} & \framebox{0} & 2 & 0 & 1 & 1 & 2 & 1 \\
& 2 & 0 & 2 & 1 & 2 & 1 & 0 & 2 & 2 & 1 & 1 & 0 \\
& 2 & 0 & 1 & 2 & 1 & 1 & 2 & 2 & 0 & 1 & 2 & 1 \\
\end{tabular}
\caption{A perfect hash family \PHF$(6; 12, 3, 3)$. \label{fig:PHF}}
\end{figure}

A perfect hash family has at least one row that separates the $t$ columns into $t$ parts in every $m \times t$ subarray.
A weaker condition separates the $t$ columns into classes.
A $\{w_1,\dots, w_s \}$-\emph{separating hash family}, denoted \SHF$(m;n,k,\{w_1,\dots, w_s \})$, with $t = \sum_{i=1}^s w_i$, is an $m \times n$ array on $k$ symbols in which for every $m \times t$ subarray, and every way to partition the $t$ columns into classes of sizes $w_1, \ldots, w_s$, there is at least one row in which no two classes contain the same symbol; see \cite{BESZ,StinsonWC08}.
A ${\cal W}$-\emph{separating hash family}, denoted \SHF$(m;n,k,{\cal W})$, is a $\{w_1,\dots, w_s \}$-separating hash family for each $\{w_1,\dots, w_s \} \in \mathcal{W}$.
Fig.~\ref{fig:SHF} gives an example of a $\{1,2\}$-separating hash family \SHF$(3; 16, 4, \{1,2\})$.
For the $3 \times 3$ subarray consisting of columns $11, 15$, and $16$,  for example, the last row separates columns $\{11,16\}$ from column $\{15\}$.

\begin{figure}[h]
\centering
\begin{tabular}{ccccccccccccccccc}
&  & & & & & & & & & & $\downarrow$ & & & & $\downarrow$ & $\downarrow$ \\
& 1 & 1 & 1 & 1 & 2 & 2 & 2 & 2 & 3 & 3 & 3 & 3 & 4 & 4 & 4 & 4 \\
& 1 & 2 & 3 & 4 & 1 & 2 & 3 & 4 & 1 & 2 & 3 & 4 & 1 & 2 & 3 & 4 \\
$\rightarrow$ & 1 & 2 & 3 & 4 & 2 & 1 & 4 & 3 & 3 & 4 & \framebox{1} & 2 & 4 & 3 & \dashbox{2} & \framebox{1} \\
\end{tabular}
\caption{A $\{1,2\}$-separating hash family \SHF$(3; 16, 4, \{1,2\})$. \label{fig:SHF}}
\end{figure}

A \emph{distributing hash family} \DHF$(m; n, k, t, s)$ is an \SHF$(m; n, k, {\cal W})$ with ${\cal W} = \{ \{ w_1,\ldots,w_s \} : t = \sum_{i=1}^s w_i \}$.
Fig.~\ref{fig:DHF} gives an example of a \DHF$(10; 13,9,5,2)$.
For the $10 \times 5$ subarray consisting of columns 8 through 12, row 4 separates columns $\{8,9,10,11\}$ from column $\{12\}$ (a $\{1,4\}$-separation), and row 5 separates columns $\{8,9,12\}$ from columns $\{10,11\}$ (a $\{2,3\}$-separation).

\begin{figure}[h!]
\centering
\begin{tabular}{cccccccccccccc}
& &  & & & & & & $\downarrow$ & $\downarrow$ & $\downarrow$ &$\downarrow$ & $\downarrow$ & \\
& 6 & 7 & 8 & 3 & 4 & 0 & 2 & 2 & 3 & 0 & 5 & 1 & 1 \\
& 3 & 1 & 1 & 7 & 2 & 6 & 8 & 4 & 3 & 0 & 2 & 0 & 5 \\
& 8 & 5 & 1 & 4 & 2 & 3 & 2 & 6 & 7 & 0 & 1 & 3 & 0 \\
a $\{1,4\}$-separation $\rightarrow$ & 0 & 2 & 0 & 2 & 2 & 0 & 0 & \framebox{1} & \framebox{1} & \framebox{1} & \framebox{1} & \dashbox{2} & 0 \\
a $\{2,3\}$-separation $\rightarrow$ & 0 & 0 & 2 & 1 & 1 & 1 & 2 & \framebox{0} & \framebox{0} & \dashbox{2} & \dashbox{2} & \framebox{0} & 1 \\
& 1 & 1 & 2 & 2 & 2 & 0 & 1 & 0 & 0 & 2 & 1 & 0 & 0 \\
& 1 & 0 & 1 & 2 & 0 & 0 & 2 & 0 & 0 & 1 & 2 & 2 & 1 \\
& 1 & 1 & 0 & 1 & 0 & 4 & 2 & 0 & 2 & 0 & 1 & 0 & 2 \\
& 0 & 0 & 3 & 0 & 1 & 0 & 0 & 2 & 4 & 0 & 0 & 1 & 0 \\
& 0 & 0 & 0 & 0 & 0 & 1 & 0 & 0 & 1 & 0 & 0 & 0 & 1 \\
\end{tabular}
\caption{A distributing hash family \DHF$(10; 13, 9, 5, 2)$. \label{fig:DHF}}
\end{figure}

Now, we are in a position to state the requirements on a pattern matrix $P$ that ensure that the sparsity supported by the matrix $B$ resulting from column replacement is at least that of $A$. \\

\begin{theorem}\label{l0infl} {\rm \cite{CHM-TCom}}
Suppose that $A$ is an $r \times k$ measurement matrix that meets the $(\ell_0,t)$-null space condition, that $P$ is an \SHF$(m;n,k,\{1,t\})$, and that $B$ is the column replacement of $A$ into $P$. Then $B$ is an $rm \times n$ measurement matrix that meets the $(\ell_0,t)$-null space condition.
\end{theorem}

\hspace*{0.25in} \\

\begin{theorem}\label{l1infl} {\rm \cite{CHM-TCom}}
Suppose that $A$ is an $r \times k$ measurement matrix that meets the $(\ell_1,t)$-null space condition, that $P$ is a \DHF$(m;n,k,t+1,2)$, and that $B$ is the column replacement of $A$ into $P$. Then $B$ is an $rm \times n$ measurement matrix that meets the $(\ell_1,t)$-null space condition.
\end{theorem}


\subsection{Exploiting Heterogeneity in Column Replacement} \label{sec:het-cr}

All the standard definitions of hash families may be generalized by replacing $k$ by $\mathbf{k} = (k_1,\dots,k_m)$, a tuple of positive integers.
Now, an \emph{heterogeneous} hash family $\HF(m; n, \mathbf{k})$, $P = (p_{ij})$, is an $m \times n$ array in which each cell from row $i$ contains one symbol from a set of $k_i$ symbols, $1 \leq i \leq m$.

Column replacement may be extended to exploit heterogeneity in an hash family.
Let $P=(p_{ij})$ be an \HF$(m;n,\mathbf{k})$ and, for $1 \leq i \leq m$, let $A^i$ be an $r_i \times k_i$ ingredient matrix whose columns are indexed by the $k_i$ elements in row $i$ of $P$.
For each row $i$ of $P$, replace the element $p_{ij}$ with a copy of column $p_{ij}$ of $A^i$, $1 \le j \le n$.
The result is a $(\sum_{i=1}^m r_i) \times n$ matrix $B$, the \emph{column replacement of $A^1,\ldots,A^m$ into $P$}.
Fig.~\ref{fig:colrep} gives an example of column replacement using an heterogeneous hash family.

\begin{figure}[h]
\begin{minipage}[b]{0.24\linewidth}
\begin{displaymath}
P = \left[ \begin{array}{c}
{\tt 132123}\\
{\tt 111222} \\
\end{array} \right]
\end{displaymath}
\end{minipage}
\begin{minipage}[b]{0.24\linewidth}
\begin{displaymath}
A^1 = \left[ \begin{array}{c}
{\tt a}^1_{11}{\tt a}^1_{12}{\tt a}^1_{13} \\
{\tt a}^1_{21}{\tt a}^1_{22}{\tt a}^1_{23} \\
\end{array} \right]
\end{displaymath}
\end{minipage}
\begin{minipage}[b]{0.24\linewidth}
\begin{displaymath}
A^2 = \left[ \begin{array}{c}
{\tt a}^2_{11}{\tt a}^2_{12} \\
{\tt a}^2_{21}{\tt a}^2_{22} \\
\end{array} \right]
\end{displaymath}
\end{minipage}
\begin{minipage}[b]{0.24\linewidth}
\begin{displaymath}
B = \left[ \begin{array}{c}
{\tt a}^1_{11}{\tt a}^1_{13}{\tt a}^1_{12}{\tt a}^1_{11}{\tt a}^1_{12}{\tt a}^1_{13} \\
{\tt a}^1_{21}{\tt a}^1_{23}{\tt a}^1_{22}{\tt a}^1_{21}{\tt a}^1_{22}{\tt a}^1_{23} \\ 
{\tt a}^2_{11}{\tt a}^2_{11}{\tt a}^2_{11}{\tt a}^2_{12}{\tt a}^2_{12}{\tt a}^2_{12} \\
{\tt a}^2_{21}{\tt a}^2_{21}{\tt a}^2_{21}{\tt a}^2_{22}{\tt a}^2_{22}{\tt a}^2_{22} \\
\end{array} \right]
\end{displaymath}
\end{minipage}
\caption{$B$ is the column replacement of $A^1, A^2$ into $P$. \label{fig:colrep}}
\end{figure}

An hierarchical method for compressive sensing is obtained using column replacement in an heterogeneous hash family.
Suppose that $A^i$ is a measurement matrix for a signal of dimension $k_i$ supporting the recovery of sparsity $q_i$, for $1 \leq i \leq m$.
We now describe the properties the pattern matrix needs to satisfy to support recovery of signals of dimension $n$ and sparsity $t$.

In Section \ref{sec:crhf-cs}, we saw that a perfect hash family separates $t$ columns into $t$ parts, and that a separating hash family separates $t$ columns into classes.
We now define a particular type of separating hash family in which the number of symbols used to accomplish the separations is restricted.

Let $\mathbf{d} = (d_1,\dots,d_m)$ be a tuple of positive integers, and let $\tau$ be a positive integer.
Let ${\cal W}  =  \{W_1, \dots, W_r\}$, where for $1 \leq i \leq r$, $W_i = \{w_{i1}, \dots, w_{is_i}\}$ is a multiset of nonnegative integers, and $\sigma_i = \sum_{j=1}^{s_i} w_{ij}$.
An $\SHF(m; n, \mathbf{k},{\cal W})$, $P = (p_{ij}),$ is $(\mathbf{d},\tau)$-\emph{strengthening} if whenever $1 \leq i \leq r$,
\begin{itemize}
	\item $C$ is a set of $\sigma_i$ columns,
	\item $C_1, \dots , C_{s_i}$ is a partition of $C$ with $|C_j| = w_{ij}$ for $1 \leq j \leq s_i$,  and
	\item $T$ is a set of $\tau$ columns with $|C \cap T| = \min(\sigma_i,\tau)$,
\end{itemize}
there exists a row $\rho$ for which $p_{\rho x} \neq p_{\rho y}$ whenever  $x \in C_e$, $y \in C_f$ and $e \neq f$ {\sl and} the multiset $\{p_{\rho x} : x \in T\}$ contains no more than $d_\rho$ different symbols.
When $\tau = \max\{\sigma_i : 1 \leq i \leq r\}$, we omit $\tau$ and write $\bf d$-strengthening.
Because rows of $P$ can be arbitrarily permuted (while permuting the ingredient matrices in the same manner), the order of elements in $\mathbf{k}$ and $\mathbf{d}$ is inconsequential.
Hence we often use exponential notation, writing $x_1^{u_1} \cdots x_s^{u_s}$, with $u_i$ a non-negative integer for $1 \leq i \leq s$, for a vector $(y_1,\dots,y_{\sum_{j=1}^s u_j})$ in which $y_\ell = x_j$ for $\sum_{j=1}^{\ell-1} u_j  < \ell \leq \sum_{j=1}^\ell u_j $ for $1\leq \ell \leq \sum_{j=1}^s u_j$.

Fig.~\ref{fig:strengthDHF} gives a heterogeneous {\bf d}-strengthening \DHF$(19;13,\mathbf{k},5,2)$ with $\mathbf{k} = (5^6 4^1 3^{12})$ and $\mathbf{d} = (4^6 3^{13})$.
This is equivalent to a {\bf d}-strengthening \SHF$(19;13,\mathbf{k},\{\{1,4\},\{2,3\}\})$.
Consider the separation of columns $\{ 1, 7\}$ from columns $\{ 2, 6, 11\}$.
Row 8 accomplishes the required separation because it uses no more than $d_8 = 3$ symbols.
Consider instead columns $\{1,\ldots,5\}$.
While the first row separates $\{1,2,3\}$ from $\{4,5\}$, it uses 5 symbols instead of $d_1 = 4$ and so does not accomplish the required separation; this separation is accomplished in row 3.

\begin{figure}[h]
\centering
\begin{tabular}{cccccccccccccc}
& $\Downarrow$ & $\downarrow$ & & & & $\downarrow$ & $\Downarrow$ & & & & $\downarrow$ & & \\
& 4 & 0 & 2 & 1 & 3 & 3 & 0 & 0 & 1 & 4 & 2 & 2 & 1 \\
& 0 & 0 & 1 & 1 & 2 & 3 & 1 & 3 & 2 & 4 & 2 & 0 & 4 \\
$k_1= \ldots= k_6 = 5$ symbols; & 0 & 2 & 4 & 1 & 1 & 2 & 0 & 1 & 2 & 3 & 0 & 3 & 4 \\
$d_1= \ldots = d_6 = 4$ used to separate & 2 & 4 & 1 & 0 & 3 & 0 & 3 & 1 & 1 & 4 & 2 & 0 & 2 \\
& 2 & 1 & 2 & 2 & 4 & 0 & 0 & 4 & 0 & 1 & 1 & 3 & 3 \\
& 3 & 4 & 0 & 1 & 0 & 3 & 2 & 4 & 2 & 1 & 1 & 2 & 0 \\ \hline
\multirow{1}{*}{$k_2=4$ symbols; $d_2=3$ used to separate}& 0 & 0 & 1 & 0 & 0 & 2 & 2 & 0 & 0 & 1 & 3 & 0 & 0 \\ \hline
$\rightarrow$ & \framebox{0} & \dashbox{1} & 0 & 1 & 1 & \dashbox{2} & \framebox{0} & 2 & 0 & 0 & \dashbox{2} & 1 & 2 \\
& 1 & 0 & 2 & 0 & 0 & 2 & 1 & 1 & 0 & 2 & 0 & 2 & 1 \\
& 0 & 1 & 2 & 2 & 1 & 2 & 1 & 0 & 0 & 0 & 1 & 0 & 2 \\
& 0 & 2 & 2 & 2 & 1 & 2 & 0 & 0 & 1 & 1 & 0 & 0 & 1 \\
& 0 & 1 & 1 & 0 & 1 & 2 & 1 & 0 & 0 & 2 & 0 & 2 & 2 \\
$k_8= \ldots= k_{19} = 3$ symbols; & 2 & 1 & 0 & 1 & 2 & 0 & 1 & 2 & 0 & 2 & 0 & 0 & 1 \\
$d_8= \ldots= d_{19} = 3$ used to separate & 2 & 1 & 2 & 0 & 2 & 2 & 0 & 0 & 1 & 0 & 0 & 1 & 1 \\
& 0 & 0 & 0 & 1 & 0 & 1 & 1 & 2 & 0 & 1 & 2 & 2 & 2 \\
& 1 & 2 & 0 & 1 & 1 & 1 & 2 & 2 & 0 & 0 & 0 & 2 & 0 \\
& 1 & 0 & 2 & 1 & 1 & 0 & 0 & 2 & 0 & 0 & 2 & 1 & 0 \\
& 2 & 2 & 0 & 0 & 1 & 2 & 1 & 0 & 0 & 1 & 2 & 1 & 2 \\
& 0 & 0 & 2 & 1 & 1 & 0 & 1 & 2 & 0 & 2 & 2 & 1 & 1 \\
\end{tabular}
\caption{A heterogeneous {\bf d}-strengthening \DHF$(19;13,\mathbf{k},5,2)$ with $\mathbf{k} = (5^6 4^1 3^{12})$ and $\mathbf{d} = (4^6 3^{13})$. \label{fig:strengthDHF}}
\end{figure}

Next the properties are determined for an heterogeneous hash family to support recovery of signals of dimension $n$ and sparsity $t$ using a column replacement technique.\\

\begin{theorem}\label{thm:l0inflst} {\rm \cite{ColbournHorsleySyrotiuk11}}
Let $\mathbf{k}=(k_1,\ldots,k_m)$ and $\mathbf{q}=(q_1,\ldots,q_m)$ be tuples of positive integers.
Let $\mathbf{d} = (2q_1,\dots,2q_m)$.
For $1 \leq i \leq m$, let $A^i  \in {\mathbb R}^{r_i \times k_i}$ be a measurement matrix that meets the $(\ell_0,q_i)$-null space condition.
Let $P$ be a $(\mathbf{d},2t)$-strengthening \SHF$(m;n,\mathbf{k},\{1,t\})$, and let $B$ be the column replacement of $A^1,\ldots,A^m$ into $P$.
Then $B$ meets the $(\ell_0,t)$-null space condition.
\end{theorem}

\hspace*{0.5in} \\

\begin{theorem}\label{thm:l1inflst} {\rm \cite{ColbournHorsleySyrotiuk11}}
Let $\mathbf{k}=(k_1,\ldots,k_m)$ and $\mathbf{q}=(q_1,\ldots,q_m)$ be tuples of positive integers.
For $1 \leq i \leq m$, let $A^i  \in {\mathbb R}^{r_i \times k_i}$ be a measurement matrix that meets the $(\ell_1,q_i)$-null space condition.
Let $P$ be a $(\mathbf{q},t)$-strengthening \DHF$(m;n,\mathbf{k},t+1,2)$, and let $B$ be the column replacement of $A^1,\ldots,A^m$ into $P$.
Then $B$ meets the $(\ell_1,t)$-null space condition.
\end{theorem}

\hspace*{0.5in} \\

Revisiting the {\bf d}-strengthening \DHF$(19;13,\mathbf{k},5,2)$ pattern matrix in Fig.~\ref{fig:strengthDHF}, the results of Theorems \ref{thm:l0inflst} and \ref{thm:l1inflst} indicate that the number of symbols in each row need not be the same.
In general, there may be as many ingredient matrices $A^i$ as there are rows of the pattern matrix $P$.
Moreover, the strength of each ingredient matrix $A^i$ may be different!
In this example, the first 6 rows use 4 symbols to separate, so the corresponding ingredient matrices must have strength at least 4.
The remaining rows use 3 symbols to separate, so the corresponding ingredient matrices must have strength at least 3.

In \cite{ColbournHorsleySyrotiuk11}, we showed that heterogeneity gives great flexibility in construction of  measurement matrices using column replacement.
The hierarchical structure of the measurement matrices produced by column replacement can also aid in recovery, and be used to support hybrid recovery schemes.
We examine this problem next, considering a generalization of hash families that removes the restriction to those strategies based only on $(\ell_0,t)$- or $(\ell_1,t)$-recoverability.
We also consider the computational investment required to recover the signal.


\section{Hash Families for Recovery} \label{sec:gen-hf}

In order to tackle signal recovery, we require another generalization of hash families.
As before, let $\mathbf{k} = (k_1,\dots,k_m)$ be a tuple of positive integers.
An $\rHF(m; n, \mathbf{k})$ is an $m \times n$ array, $P = (p_{ij})$, in which each cell contains one symbol, and for each row $1 \leq i \leq m$, $\{p_{ij} : 1 \leq j \leq n\} \subseteq \{\circ,1,\ldots,k_i\}$.
The symbol $\circ$, when present, is interpreted as representing a `missing' entry.
When the pattern matrix $P=(p_{ij})$ is an \rHF$(m;n,\mathbf{k})$,  and for $1 \leq i \leq m$ the ingredient matrix $A^i$ is $r_i \times k_i$ with columns indexed by the $k_i$ symbols in row $i$ of $P$ other than $\circ$, the column replacement of $A^1,\dots,A^m$ into $P$ is as before, except that when $p_{ij} = \circ$, it is replaced with an all zero column vector of length $r_i$.
As we will see, the separating properties of the hash families we use allow us to locate the nonzero coordinates of the signal and hence perform the recovery.

The definition of a $\mathcal W$-separating hash family encompasses perfect, $\{w_1, \ldots, w_s\}$-separating, and distributing hash families.
Therefore, we need only extend the definition of  $\mathcal W$-separating hash families to include the $\circ$ symbol.
To do so, we allow some of the elements of the multisets in $\mathcal W$ to be marked with a $\circ$ superscript to form a set of
marked multisets $\mathcal W'$; the multisets in $\mathcal W'$ are indexed.
Then an \rHF$(m;n,{\bf k})$ is $\mathcal W'$-separating if, for each $\{w_1, \ldots, w_s \} \in \mathcal W$ (with some elements possibly marked),
\begin{itemize}
	\item whenever $C$ is a set of $\sum_{i=1}^s w_i$ columns, and
	\item $C_1, \dots, C_s$ is an (indexed) partition of $C$ with $|C_i| = w_i$ for $1 \leq i \leq s$
\end{itemize}
then there exists a row that separates $C_1,\ldots,C_s$ in which, for $1 \leq j \leq s$, if $\circ$ appears in a column in $C_j$  then $w_j$ is marked.

As we will see, to recover the signal, the idea is to effect a separation where a significant coordinate of the signal is present in one class such that any other class does not prevent its recovery.


\subsection{Signal Recovery without Noise} \label{sec:recovery-nonoise}

Theorems \ref{thm:l0inflst} and \ref{thm:l1inflst} suggest that a recovery scheme based on $(\ell_0,t)$- or $(\ell_1,t)$-recoverability can be `lifted' from the ingredient measurement matrices $A^1,\dots,A^m$ to the larger measurement matrix $B$ obtained from column replacement.
However, such a method appears to have two main drawbacks.
First, it is restricted to recovery strategies based on $(\ell_0,t)$- or $(\ell_1,t)$-recoverability.
Secondly, and perhaps more importantly, it appears to necessitate a large computational investment to recover the signal, given $B$.

In order to overcome these problems, we consider two cases.
The \emph{positive} case arises when the signal is known {\sl a priori} to be in ${\mathbb R}_{\geq 0}^n$.
The \emph{general} case arises when the signal can be positive, negative, or zero.
In each case we develop a recovery scheme for the matrix $B$ resulting from column replacement that does not depend on any fixed algorithm, but rather on the recovery schemes for the ingredient matrices $A^1,\dots,A^m$.

We suppose that $P=(p_{ij})$ is an \rHF$(m;n,{\bf k})$.
For each $1 \leq i \leq m$, we suppose that $A^i$ is an $r_i \times k_i$ measurement matrix that has $(\ell_0,t)$-recoverability, equipped with a recovery algorithm ${\cal R}_i$ that determines the unique $t$-sparse vector ${\bf z}_i$ that solves $A^i {\bf z}_i = {\bf y}_i$.
We further suppose that $B$ is the column replacement of $A^1,\dots,A^m$ into $P$, and that $\mathbf{y}$ is the result of sampling an (unknown) $t$-sparse vector $\mathbf{x}=(x_1,\ldots,x_n)$ using $B$.

For $1 \leq i \leq m$, the $i$th row of $P$ induces a partition $\{S_{i\circ},S_{i1},\ldots,S_{ik_i}\}$ of the column indices $\{1,\dots,n\}$, where $S_{i\sigma}=\{j: p_{ij} = \sigma, 1 \leq j \leq n\}$ for $\sigma \in \{\circ,1,\ldots,k_i\}$.
Assume that we have employed the recovery algorithms ${\cal R}_i$ to find solutions ${\bf z}_i$.
For $1 \leq i \leq m$ and $\sigma \in \{\circ,1,\ldots,k_i\}$, the partition class $S_{i\sigma}$ is \emph{discarded} if $\sigma=\circ$, \emph{insignificant} if $\sigma \neq \circ$ and $z_{i\sigma}=0$, \emph{significant positive} if $z_{i\sigma}>0$, and \emph{significant negative} if $z_{i\sigma}<0$.

For $1 \leq i \leq m$, let $\mathbf{w_i}=(w_{i1},\ldots,w_{ik_i})$ where $w_{i\sigma}=\sum_{j \in S_{i\sigma}}x_j$.
The vector $\mathbf{w_i}$ can be considered as a projection of $\mathbf{x}$ induced by the symbol pattern in row $i$ of $P$.
These facts follow:
\begin{itemize}
	\item For $1 \leq i \leq m$, by the definition of $B$ and because $B_i\mathbf{x}=\mathbf{y_i}$, $\mathbf{z_i}=\mathbf{w_i}$ is a solution to $A^i\mathbf{z_i}=\mathbf{y_i}$.

	\item For $1 \leq i \leq m$, because $A^i$ has $(\ell_0,t)$-recoverability and $\mathbf{w_i}$ is $t$-sparse (because $\mathbf{x}$ is $t$-sparse), $\mathbf{z_i}=\mathbf{w_i}$ is the unique solution to $A^i\mathbf{z_i}=\mathbf{y_i}$, and so ${\cal R}_i$ returns $\mathbf{w_i}$.
\end{itemize}

We now consider the positive case and the general case for recovery in succession.


\subsection{Signal Recovery: The Positive Case} \label{sec:positive-case}

We establish that in the positive case with $t$-sparse signals, it suffices to use a separating hash family of suitable strength, along with suitable ingredient matrices.
An \rSHF$(m;n,\mathbf{k},\{1,t^\circ\})$ separates $t+1$ columns into two parts, one part of size one that cannot include the symbol $\circ$, and the other of size $t$ that may include $\circ$. \\

\begin{theorem}\label{thm:posnonoise}
Suppose that $P$ is an \rSHF$(m;n,\mathbf{k},\{1,t^\circ\})$.
For $1 \leq i \leq m$, let $A^i  \in {\mathbb R}^{r_i \times k_i}$ be a measurement matrix that has $(\ell_0,t)$-recoverability equipped with a recovery algorithm ${\mathcal R}_i$ that determines the unique $t$-sparse vector ${\bf z}_i$ that solves $A^i {\bf z}_i = {\bf y}_i$.
Further suppose that $B$ is the column replacement of measurement matrices $A^1,\ldots,A^m$ into $P$ and that $\mathbf{y}$ is the result of sampling an (unknown) $t$-sparse vector $\mathbf{x}=(x_1,\ldots,x_n) \in {\mathbb R}_{\geq 0}^n$ using $B$.
Then the $t$-sparse solution $\bf x$ to $B {\bf x} = {\bf y}$ can be recovered.
\end{theorem}

\hspace*{0.5in} \\

\begin{IEEEproof}
It suffices to determine whether $x_i$ is positive or zero for each $1 \leq i \leq n$, because once this is accomplished we can find the values of the positive $x_i$ by solving the overdetermined system that remains.
For $1 \leq i \leq m$, apply recovery algorithm ${\mathcal R}_i$ to find the unique $t$-sparse vector ${\bf z}_i$ such that $A^i {\bf z}_i = {\bf y}_i$.
We claim that, for $1 \leq \ell \leq n$, $x_\ell$ is positive if and only if for each $i \in \{1,\ldots,m\}$ the partition class that contains $\ell$ is either significant positive or discarded.

Suppose first that $x_\ell$ is positive.
If $S_{i\sigma}$ is a partition class that contains $\ell$, then either $\sigma=\circ$ and $S_{i\sigma}$ is insignificant or $\sigma \neq \circ$, $z_{i\sigma}=\sum_{j\in S_{i\sigma}}x_j \geq x_\ell >0$, and $S_{i\sigma}$ is significant positive.
Now suppose that $x_\ell=0$. Let $C=\{j:x_j >0,1 \leq j \leq n\}$; $|C|\leq t$.
There must be a row $\rho$ of $P$ that separates $C$ from $\{\ell\}$ such that $p_{\rho \ell} \neq \circ$.
Let $\sigma=p_{\rho \ell}$.
Then $\ell \in S_{\rho\sigma}$ and $S_{\rho\sigma} \cap C = \emptyset$, so $S_{\rho\sigma}$ is insignificant.
\end{IEEEproof}

\hspace*{0.5in} \\

One useful application of Theorem \ref{thm:posnonoise} takes the pattern matrix $P$ to be an \rSHF$(m;n,1,\{1,t^\circ\})$, and each $A^i$ to be a $1 \times 1$ matrix whose only element is $1$; in this case, column replacement yields a matrix $B$ isomorphic to $P$.
In $P$ for every column $\gamma$ and every set $C$ of $t$ columns with $\gamma \not\in C$, there is a row in which all columns of $C$ contain $\circ$, while column $\gamma$ contains $1$.
Then the measurement matrices $A^i$ have $(\ell_0,t)$-recoverability and the recovery algorithms $\mathcal{R}_i$ are trivial.
Hence in these cases, a matrix isomorphic to $P$ itself supports recovery.

Theorem \ref{thm:posnonoise} leads to a straightforward recovery algorithm.
First, ${\cal R}_i$ is used to solve $A^i {\bf z}_i = {\bf y}_i$ for $1 \leq i \leq m$.
Then the classes $S_{ij}$ are classified as positive when $z_{ij} > 0$, discarded when $j = \circ$, and insignificant when $j \neq \circ$ and $z_{ij} = 0$; this can be done in $\mbox{O}(\sum_{i=1}^m k_i)$ time. 
We need only compute, for each row, the complement of the union of the insignificant classes, and then compute the intersection over  all rows of these complements.
However, without additional structure this appears to require the examination of each coordinate; hence, this gives an $\Omega(n)$ lower bound.

It is not difficult, nevertheless, to obtain sublinear recovery times by restricting the hash family; we return to this problem in Section \ref{sec:sublinear}.


\subsection{Signal Recovery: The General Case} \label{sec:general-case}

When the signal takes on both positive and negative values, cancellation of positive and negative contributions can yield a zero measurement despite the presence of a signal.
Nevertheless, an additional requirement on the structure of the hash family suffices to address this problem, as we show next. \\

\begin{theorem}\label{thm:gennonoise}
Suppose that $P$ is an \rSHF$(m;n,\mathbf{k},\{\{\tau,(t+1-\tau)^\circ\} : 1 \leq \tau \leq t\})$.
For $1 \leq i \leq m$, let $A^i  \in {\mathbb R}^{r_i \times k_i}$ be a measurement matrix that has $(\ell_0,t)$-recoverability equipped with a recovery algorithm ${\mathcal R}_i$ that determines the unique $t$-sparse vector ${\bf z}_i$ that solves $A^i {\bf z}_i = {\bf y}_i$.
Further suppose that $B$ is the column replacement of measurement matrices $A^1,\ldots,A^m$ into $P$ and that $\mathbf{y}$ is the result of sampling an (unknown) $t$-sparse vector $\mathbf{x}=(x_1,\ldots,x_n)$ using $B$.
Then the $t$-sparse solution $\bf x$ to $B {\bf x} = {\bf y}$ can be recovered.
\end{theorem}

\hspace*{0.5in} \\

\begin{IEEEproof}
As in the proof of Theorem~\ref{thm:posnonoise}, it suffices to determine whether $x_i$ is nonzero or zero for each $1 \leq i \leq n$, because once this is accomplished we can find the values of the nonzero $x_i$ by solving the overdetermined system that remains.
For $1 \leq i \leq m$, apply recovery algorithm ${\mathcal R}_i$ to find the unique $t$-sparse vector ${\bf z}_i$ such that $A^i {\bf z}_i = {\bf y}_i$.
Let ${\bf z}_i^+ = (\max(0,z_{ij}) : 1 \leq j \leq k_i)$ and ${\bf z}_i^- = (\min(0,z_{ij}) : 1 \leq j \leq k_i)$.

A row $i$ of $P$ is \emph{maximum positive} if $|| {\bf z}_i^+ ||_1 \geq || {\bf z}_{i'}^+ ||_1$ for $1 \leq i' \leq m$.
Let $M \subseteq \{1,\dots, m\}$ index the maximum positive rows.
We claim that a coordinate $x_\ell$ is positive if and only if, for every $\rho \in M$, $\ell$ is in a significant positive class of the partition induced by row $\rho$.

Suppose first that $x_\ell$ is positive and let $\rho \in M$.
Because $\rho$ indexes a maximum positive row, the partition class induced by row $\rho$ that contains $\ell$ is not discarded and does not contain the index of any negative variable.
Thus it is in a significant positive partition class.

Now suppose that $x_\ell \leq 0$.
Because $P$ is an \rSHF$(m;n,\mathbf{k},\{\{\tau,(t+1-\tau)^\circ\} : 1 \leq \tau \leq t\})$ and $\mathbf{x}$ is $t$-sparse, there is a row $\rho$ of $P$ that separates $\{j : x_j > 0, 1 \leq j \leq n\}$ from $\{j : x_j < 0, 1 \leq j \leq n\} \cup \{\ell\}$ in which the symbol $\circ$ only appears in a subset of the columns indexed by $\{j : x_j < 0, 1 \leq j \leq n\} \cup \{\ell\}$.
It follows that $\rho$ is a maximum positive row of $P$ and that the partition class induced by $\rho$ containing $\ell$
does not contain the index of any positive coordinate.
So $\rho \in M$, but $\ell$ is not in a significant positive class of the partition induced by row $\rho$.

In the same manner, all negative coordinates can be identified using maximum negative rows.
\end{IEEEproof}

\hspace*{0.5in} \\

Again a straightforward recovery algorithm is given by Theorem \ref{thm:gennonoise} but, as in the positive case, it naively involves examining each of the $n$ coordinates.


\subsection{Sublinear Time Signal Recovery} \label{sec:sublinear}

Recovery can be accomplished in time that is sublinear in $k$ when the hash family has suitable structure; we develop a general approach, and one example, here.
In each case, for some subset $M$ of the rows of $P$, sets are identified that must contain the indices of all positive coordinates (the indices of the negative coordinates, if they exist, can be located similarly).
Recall from Section \ref{sec:recovery-nonoise},  that the \emph{positive} case arises when the signal is known {\sl a priori} to be in ${\mathbb R}_{\geq 0}^n$ and the \emph{general} case arises when the signal can be positive, negative, or zero.
In the positive case,  $M$ contains all rows and for $\rho \in M$, the candidate indices are $V_\rho^+ = \{ \ell : p_{\rho \ell} = \circ \mbox{, or } x_\ell \in S_{\rho j} \mbox{ and } z_{\rho j} > 0\}$.
In the general case, $M$ contains all rows that index maximum positive rows, and for $\rho \in M$, the candidate indices are $V_\rho^+ = \{ \ell : x_\ell \in S_{\rho j} \mbox{ and } z_{\rho j} > 0\}$.
In both cases, we are to determine $\bigcap_{\rho \in M} V_\rho^+$.
In order to avoid the examination of each coordinate, we do not list the members of $V_\rho^+$ explicitly, but rather use an implicit representation to list the members of $\bigcap_{\rho \in M} V_\rho^+$.

First we give an implicit representation of an hash family \HF$(q+1;q^\alpha,q)$, $P$, where $q$ is a prime power and $2 \leq \alpha \leq q$.
Let $\{\omega_0, \dots, \omega_{q-1}\}$ be the elements of the finite field of order $q$, ${\mathbb F}_q$.
Index the rows of $P$ by $\{\infty\} \cup \{\omega_0, \dots, \omega_{q-1}\}$.
Index the columns of $P$ by the $q^\alpha$ polynomials of degree less than $\alpha$ in indeterminate $x$, with coefficients in ${\mathbb F}_q$.
Now the entry of $P$ with row index $\beta$ and column indexed by polynomial $f(x)$ is determined as $f(\beta)$ when $\beta \in \{\omega_0, \dots, \omega_{q-1}\}$, and as the coefficient of $x^{\alpha-1}$ in $f(x)$ when $\beta = \infty$.

By deleting rows, we form an \HF$(m;q^\alpha,q)$ for some $1 \leq m \leq q+1$.
An hash family is \emph{linear} if it is obtained in this way.
The separation properties of such an hash family are crucial \cite{Alon86,BlackburnWild}.
For our purposes, the observation of interest is from \cite{CLlinhf}: if $m \geq (\alpha-1)w_1w_2+1$, then a linear \HF$(m;n,q)$ is $\{w_1,w_2\}$-separating.
(This can be established by a simple argument: When two polynomials of degree less than $\alpha$ evaluate to the same value at $\alpha$ different points, they are the same polynomial.)
In some cases, fewer rows suffice to ensure separation.
In particular, Blackburn and Wild \cite{BlackburnWild} establish that when $q$ is sufficiently large, one needs at most $\alpha(w_1+w_2-1)$ rows; and in \cite{CLlinhf} specific small separations are examined to determine the set of prime powers for which various numbers of rows less than $(\alpha-1)w_1w_2+1$ suffice.
We proceed with the general statement so as not to impose additional conditions.

When $m \geq (\alpha-1)t+1$, $P$  is $\{1,t\}$-separating; in addition, every $\{1,t-1\}$-separation is accomplished in at least $\alpha$ rows.
When $m \geq (\alpha-1) \lfloor \frac{t+1}{2} \rfloor \lceil \frac{t+1}{2} \rceil +1$, $P$ is $\{w,t+1-w\}$-separating for each $1 \leq w \leq t$; in addition, every $\{w,t-w\}$-separation is accomplished in at least $\alpha$ rows, because $\lfloor \frac{t+1}{2} \rfloor \lceil \frac{t+1}{2} \rceil = \lfloor \frac{t}{2} \rfloor \lceil \frac{t}{2} \rceil + \lfloor \frac{t+1}{2} \rfloor$.
Thus in either case, $M$ contains at least $\alpha$ rows of $P$.

Choose any $\alpha$ rows  $U = \{\psi_1, \dots \psi_{\alpha}\} \subseteq M$.
Now consider the sets $\{ V^+_\psi : \psi \in U\}$.
Define $\prod_{\psi \in U} | V^+_\psi |$ vectors ${\mathcal V}^+ = \{ (g_1, \dots, g_{\alpha}) : g_i \in \{p_{\psi_i\ell}: \ell \in V^+_{\psi_i}\}
\mbox{ for } 1\leq i \leq \alpha\}$.
Each $(g_1, \dots, g_{\alpha}) \in {\mathcal V}^+$ defines a unique column of the hash family, corresponding to the unique polynomial $L$ of degree at most $\alpha -1$ satisfying $L(\psi_i) = g_i$ for $1 \leq i \leq \alpha$.
Any column that does {\sl not} arise in this way from a member of ${\mathcal V}^+$ {\sl cannot} be the column for a positive coordinate, because in the partition induced by one of the selected maximum rows it is not in a significant positive class.
However, columns arising from vectors in ${\mathcal V}^+$ need not arise from positive coordinates, because we may not have examined all of the rows of $M$.
Nevertheless, we can now generate each of the columns arising from vectors in ${\mathcal V}^+$, and check for each whether it occurs in positive classes for all rows of $M$, not just the $\alpha$ selected.

Now $|{\mathcal V}^+|$ is $O(t^\alpha)$, so when $t$ is $\mbox{o}(q)$, the size of ${\mathcal V}^+$ is $\mbox{o}(n)$ (because $n = q^\alpha$).
For concreteness, taking $q=t^\beta$ for $t$ a prime power, we can permit $\alpha$ to be as large as $t^{\beta-2}$.
(For the positive case, we can permit $\alpha$ to be as large as $t^{\beta-1}$. )
Hence, by restricting the hash family to one that is linear, it is possible to obtain recovery of the signal in sublinear time.

In general, a hash family together with its ingredient matrices can be represented more concisely compared to a random measurement matrix for signal recovery.
Furthermore, the hash family is an integer matrix, not a matrix of real numbers, and may therefore be easier to encode.
When the hash family is linear an implicit representation of it may be used, further compacting its representation.

The results of this section provide some evidence that column replacement enables recoverability conditions to be met.
In Section \ref{sec:noise}, we show that it also preserves the basic machinery to deal with noise in the signal.


\subsection{Adding Strengthening} \label{sec:strengthening}

As the signal length increases, it is natural to support high sparsity.
Yet the techniques developed until this point only preserve sparsity.
Strengthening hash families provide a means to increase the level of sparsity supported. \\

\begin{theorem}\label{thm:gennonoisestrengthen}
Suppose that $P$ is a $\mathbf{d}$-strengthening \SHF$(m;n,\mathbf{k},\{\{\tau,(t+1-\tau)\} : 1 \leq \tau \leq t\})$.
For each $1 \leq i \leq m$, we suppose that $A^i$ is an $r_i \times k_i$ measurement matrix that has $(\ell_1,d_i)$-recoverability, equipped with a recovery algorithm ${\cal R}_i$, that either determines the unique $d_i$-sparse vector ${\bf z}_i$ that solves $A^i {\bf z}_i = {\bf y}_i$ or indicates that no such vector exists.
Further suppose that $B$ is the column replacement of $A^1,\dots,A^m$ into $P$, and that $\mathbf{y}$ is the result of sampling an (unknown) $t$-sparse vector $\mathbf{x}=(x_1,\ldots,x_n)$ using $B$.
Then the $t$-sparse solution $\bf x$ to $B {\bf x} = {\bf y}$ can be recovered.
\end{theorem}

\hspace*{0.5in} \\

\begin{IEEEproof}
Again it suffices to locate the nonzero coordinates of $\mathbf{x}$.
For $1 \leq i \leq m$, if recovery algorithm $\mathcal{R}_i$ returns a solution $\mathbf{z_i}$ such that $||\mathbf{z_i}||_1 \geq ||\mathbf{z}||_1$ for any solution $\mathbf{z}$ returned by an oracle $\mathcal{R}_j$, then $\mathbf{z_i}$ is a \emph{maximum solution}, and row $i$ of $P$ is a \emph{maximum row}.
Because $P$ is a $\mathbf{d}$-strengthening \SHF$(m;n,\mathbf{k},\{\{\tau,(t+1-\tau)\} : 1 \leq \tau \leq t\})$, and $\mathbf{x}$ is $t$-sparse, there is a row $\rho$ of $P$ that separates $\{j : x_j > 0, 1 \leq j \leq n\}$ from $\{j : x_j < 0, 1 \leq j \leq n\}$ with the property that at most $d_\rho$ symbols appear in the columns indexed by $\{j : x_j \neq 0, 1 \leq j \leq n\}$.
So the projected vector $\mathbf{w}_\rho$ is $d_\rho$-sparse and it is the solution returned by $\mathcal{R}_\rho$.
By the definition of $\rho$, $||\mathbf{w}_\rho||_1=||\mathbf{x}||_1$.
It follows that the $\ell_1$-norm of any maximum solution is at least $||\mathbf{x}||_1$.

We claim that if $\mathcal{R}_i$ returns a maximum solution $\mathbf{z_i}$, then $\mathbf{z_i} = \mathbf{w_i}$.
Suppose otherwise.
Then, because $\mathbf{z_i}$ is a maximum solution, we have $||\mathbf{z_i}||_1 \geq ||\mathbf{x}||_1$.
Further, it is clear from the definition of $||\mathbf{w_i}||$ that $||\mathbf{w_i}||_1 \leq ||\mathbf{x}||_1$.
Thus $A^i\mathbf{z_i}=A^i\mathbf{w_i}$, $\mathbf{z_i}$ is $d_i$-sparse, and $||\mathbf{z_i}||_1 \geq ||\mathbf{w_i}||_1$, which is a contradiction to the fact that $A^i$ has $(\ell_1,d_i)$-recoverability.

Having established our claim, we can now use arguments similar to those used in the proof of Theorem \ref{thm:gennonoise} to show that a coordinate $x_\ell$ is positive (negative) if and only if, for every maximum row in $P$, $\ell$ is in a significant positive (significant negative) class of the partition induced by that row.
\end{IEEEproof}


\section{Recovery with Noise} \label{sec:noise}

We now treat the recovery of signals with noise.
A signal $(x_1,\ldots,x_n)$ is \emph{$(s,t)$-almost sparse} if there is a set $T$ of at most $t$ coordinate indices such that $\sum_{i \in \{1\ldots,n\} \setminus T}|x_i| < s$.  \\

\begin{theorem} \label{thm:gennoise}
Suppose that $P$ is an $\SHF(m;n,\mathbf{k},\{\{\tau,(t+1-\tau)\} : 1 \leq \tau \leq t\})$.
For each $1 \leq i \leq m$, we suppose that $A^i$ is an $r_i \times k_i$ measurement matrix, equipped with recovery algorithm ${\cal R}_i$, which, when applied to the sample obtained from an $(s,t)$-almost sparse signal ${\bf x}_i$, returns a vector ${\bf z}_i$ such that $||{\bf z}_i-{\bf x}_i||_1<\epsilon$.
Further suppose that $B$ is the column replacement of $A^1,\dots,A^m$ into $P$, and that $\mathbf{y}$ is the result of sampling an (unknown) $(s,t)$-almost sparse vector $\mathbf{x}=(x_1,\ldots,x_n)$ using $B$.
Then, a (perfectly) $t$-sparse vector ${\bf x}^*=(x^*_1,\ldots,x^*_n)$ such that for $1 \leq i \leq n$, $|x_i| < 2(s+\epsilon)$ if $x^*_i=0$, and $|x_i - x^*_i| < s+\epsilon$ if $x^*_i > 0$,  and such that $B {\bf x}^* = {\bf y}$, can be recovered.
\end{theorem}

\hspace*{0.5in} \\

\begin{IEEEproof}
We provide a sketch first, and then the details.
The idea is to write each coordinate of ${\bf z}$ as a sum of the signal coordinates in $T$ that contribute to it, and of a noise term $e$ that includes both the small contributions from coordinates outside $T$ and the error less than $\epsilon$ from the recovery algorithm.
For each row $\rho$ of $P$, we then split this sum into two parts: one part containing terms with the same sign as the ${\bf z}$ coordinate to which they contribute (indexed by sets $T'_{\rho}$ and $E'_{\rho}$), and another part containing terms with the opposite sign to the ${\bf z}$ coordinate to which they contribute (indexed by sets $T''_{\rho}$ and $E''_{\rho}$).
The key observation is that the sum of the terms with indices in $T''_{\rho}$ can be approximated by $\frac{1}{2}(||{\bf x}||-||{\bf z}_{\rho}||)$ and hence by $\frac{1}{2}(q-||{\bf z}_{\rho}||)$ because if $T''_{\rho}$ is empty then ${\bf z}_{\rho}$ has norm close to $||\mathbf{x}||$, and every term with index in $T''_{\rho}$ reduces $||{\bf z}_{\rho}||$.

Let $T$ be a set of at most $t$ coordinate indices such that $\sum_{i \in \{1\ldots,n\} \setminus T}|x_i| < s$.
Let $T^+=\{i \in T:x_i \geq 0\}$, $T^-=\{i \in T:x_i < 0\}$ and $q^{\dag}=\sum_{i \in T}|x_i|$.
For $1 \leq i \leq m$, apply ${\mathcal R}_i$ to ${\bf y}_i$ to find a vector ${\bf z}_i$ such that $||{\bf z}_i-{\bf w}_i||_1<\epsilon$.
For $i \in \{1,\ldots,m\}$, call $||{\bf z}_i||_1$ the \emph{signature} of row $i$ of $P$ and let $q$ be the maximum signature of any row of $P$.

For $1 \leq i \leq n$, we calculate upper and lower estimates $u(i)$ and $\ell(i)$ for $x_i$.
For each row index $\rho \in \{1,\ldots,m\}$ and each symbol $\sigma \in \{1,\ldots,k_{\rho}\}$ we define $u_{\rho\sigma}$ and $\ell_{\rho\sigma}$ as follows.
\begin{itemize}
    \item If $z_{\rho\sigma} \geq 0$, then $u_{\rho\sigma}=|z_{\rho\sigma}|+\frac{1}{2}(q-||{\bf z}_\rho||_1)$ and $\ell_{\rho\sigma}=-\frac{1}{2}(q-||{\bf z}_\rho||_1)$.
    \item If $z_{\rho\sigma} <0$, then $u_{\rho\sigma}=\frac{1}{2}(q-||{\bf z}_\rho||_1)$ and $\ell_{\rho\sigma}=-|z_{\rho\sigma}|-\frac{1}{2}(q-||{\bf z}_\rho||_1)$.
\end{itemize}
For each $i \in \{1,\ldots,n\}$ define $u_\rho(i)=u_{\rho\pi}$ and $\ell_\rho(i)=\ell_{\rho\pi}$, where $\pi$ is the symbol in row $\rho$ of $P$ such that $i \in S_{\rho\pi}$, and define $u(i)=\min\{u_{\rho}(i): 1 \leq \rho \leq m\}$ and $\ell(i)=\max\{\ell_{\rho}(i): 1 \leq \rho \leq m\}$.
By first examining a row of maximum signature, we can immediately conclude for each $i \in \{1,\ldots,n\}$ either that $u(i)=0$ or that $\ell(i)=0$.
Define a vector ${\bf x}^*=(x^*_1,\ldots,x^*_n)$ by setting $x^*_i=0$ if $|u_i|,|\ell_i| \leq s+\epsilon$, and otherwise setting $x^*_i$ equal to whichever of $u(i)$ or $\ell(i)$ has the greater absolute value.
We claim that ${\bf x}^*$ satisfies the required conditions.

To establish this claim we prove that, for $1 \leq j \leq n$,
\begin{itemize}
    \item[(i)] for each $\rho \in \{1,\ldots,m\}$, $\ell_{\rho}(j)-(s+\epsilon) < x_j < u_{\rho}(j)+(s+\epsilon)$;
    \item[(ii)] there is some $\rho \in \{1,\ldots,m\}$ such that $\ell_{\rho}(j) > -(s+\epsilon)$ if $x_j \geq 0$ and $u_{\rho}(j) < s+\epsilon$ if $x_j < 0$; and
    \item[(iii)] there is some $\rho \in \{1,\ldots,m\}$ such that $u_{\rho}(j)-(s+\epsilon) < x_j$ if $x_j \geq 0$ and $x_j < \ell_{\rho}(j)+(s+\epsilon)$ if $x_j < 0$.
\end{itemize}

We begin with some observations used throughout the proof.
Let $\rho$ be a row of $P$.
For $1 \leq \sigma \leq k_{\rho}$, we have $z_{\rho\sigma}=(\sum_{i \in T \cap S_{\rho\sigma}}|x_i|)+e_{\rho\sigma}$ for some $e_{\rho\sigma}$.
Note that $\sum_{\sigma=1}^{k_\rho}|e_{\rho\sigma}| \leq s+\epsilon$.
Let $T'_{\rho}=\{i \in T^+:z_{\rho p_{\rho i}} \geq 0\} \cup \{i \in T^-:z_{\rho p_{\rho i}} < 0\}$ and let $T''_{\rho} =T \setminus T'_{\rho}$.
Further, let $E'_\rho=\{\sigma \in \{1,\ldots,k_\rho\}:e_{\rho\sigma},z_{\rho\sigma} \geq 0 \mbox{ or } e_{\rho\sigma},z_{\rho\sigma} < 0\}$ and let $E''_{\rho} = \{1,\ldots,k_\rho\} \setminus E'_{\rho}$. For $1 \leq \pi \leq k_\rho$, we have that
\begin{equation}\label{zClassMagEq}
    |z_{\rho\pi}|=\left(\sum_{i \in T'_{\rho} \cap S_{\rho\pi}}|x_i|\right)-\left(\sum_{i \in T''_{\rho} \cap S_{\rho\pi}}|x_i|\right)+\delta_{\rho\pi} |e_{\rho\pi}|
\end{equation}
where $\delta_{\rho\pi}=1$ if $\pi \in E'_{\rho}$ and $\delta_{\rho\pi}=-1$ if $\pi \in E''_{\rho}$.
Summing over the symbols in row $\rho$ of $P$, we see
\begin{equation}\label{zNormEq}
    ||{\bf z}_{\rho}||_1=q^{\dag}-2\left(\sum_{i \in T_\rho''}|x_i|\right)+\left(\sum_{\sigma \in E_\rho'} |e_{\rho\sigma}|\right)-\left(\sum_{\sigma \in E_\rho''}|e_{\rho\sigma}|\right)
\end{equation}
and it follows that
\begin{equation}\label{MisplacedEq}
    \tfrac{1}{2}(q^{\dag}-||{\bf z}_{\rho}||_1)=\left(\sum_{i \in T_\rho''}|x_i|\right)-\frac{1}{2}\left(\sum_{\sigma \in E_\rho'} |e_{\rho\sigma}|\right)+\frac{1}{2}\left(\sum_{\sigma \in E_\rho''}|e_{\rho\sigma}|\right).
\end{equation}
Adding (\ref{zClassMagEq}) to (\ref{MisplacedEq}), we obtain
\begin{equation}\label{BigBound}
    |z_{\rho\pi}|+\tfrac{1}{2}(q^{\dag}-||{\bf z}_{\rho}||_1)=\left(\sum_{i \in T'_{\rho} \cap S_{\rho\pi}}|x_i|\right)+\left(\sum_{i \in T_\rho'' \setminus S_{\rho\pi}}|x_i|\right)-\frac{1}{2}\left(\sum_{\sigma \in E_\rho'} |e_{\rho\sigma}|\right)+\frac{1}{2}\left(\sum_{\sigma \in E_\rho''}|e_{\rho\sigma}|\right)+\delta_{\rho\pi} |e_{\rho\pi}|.
\end{equation}
It follows from (\ref{zNormEq}) that each row of $P$ has signature less than $q^{\dag}+(s+\epsilon)$ and that any row of $P$ that separates $T^+$ from $T^-$ has signature greater than $q^{\dag}-(s+\epsilon)$.
Thus, $q^{\dag}-(s+\epsilon) < q < q^{\dag}+(s+\epsilon)$ and hence
\begin{equation}\label{qBoundEq}
    \tfrac{1}{2}(q- ||{\bf z}_{\rho}||_1)-\tfrac{1}{2}(s+\epsilon) < \tfrac{1}{2}(q^{\dag} - ||{\bf z}_{\rho}||_1)
    < \tfrac{1}{2}(q- ||{\bf z}_{\rho}||_1)+\tfrac{1}{2}(s+\epsilon).
\end{equation}

Let $j \in \{1,\ldots,n\}$.
We next show that (i), (ii) and (iii) hold in the case where $x_j \geq 0$.
The proof in the case where $x_j < 0$ is similar.

\emph{Proof of (i).}
Let $\rho$ index any row of $P$ and let $S_{\rho\pi}$ be the partition class induced by row $\rho$ of $P$ that contains $j$.
Now $\ell_\rho(j)-(s+\epsilon) < x_j$ because $\ell_\rho(j) \leq 0$.
If $j \notin T$, then $x_j < s$ and  $x_j < u_{\rho}(j)+(s+\epsilon)$ because $u_\rho(j) \geq 0$.
If $j \in T$ and $z_{\rho\pi} < 0$, then $x_j \leq \sum_{i \in T_\rho''}|x_i|$ and we see from (\ref{MisplacedEq}) and (\ref{qBoundEq}) that $x_j < \frac{1}{2}(q-||{\bf z}_\rho||_1)+(s+\epsilon)$.
If $j \in T$ and $z_{\rho\pi} \geq 0$, then $x_j \leq \sum_{i \in T_\rho' \cap S_{\rho\pi}}|x_i|$ and we see from (\ref{BigBound}) and (\ref{qBoundEq}) that $x_j < |z_{\rho\pi}|+\frac{1}{2}(q-||{\bf z}_\rho||_1)+(s+\epsilon)$.

\emph{Proof of (ii).}
Let $\rho$ index a row of $P$ that separates $T^+ \cup \{j\}$ from $T^-$ and let $S_{\rho\pi}$ be the partition class induced by row $\rho$ of $P$ that contains $j$.
If $z_{\rho\pi} \geq 0$, then, for $1 \leq \sigma \leq k_\rho$, either $\sum_{i \in T''_\rho \cap S_{\rho\sigma}}|x_i| \leq |e_{\rho\sigma}|$ and $\sigma \in E'_\rho$ or $T''_{\rho} \cap S_{\rho\sigma}=\emptyset$.
Using this, it follows from (\ref{MisplacedEq}) and (\ref{qBoundEq}) that $\ell_{\rho}(j)=-\frac{1}{2}(q-||{\bf z}_\rho||_1) > -(s+\epsilon)$.
If $z_{\rho\pi} < 0$, then $T'_{\rho} \cap S_{\rho\pi}=\emptyset$ and $\pi \in E'_{\rho}$.
Furthermore, for $\sigma \in \{1,\ldots,k_\rho\} \setminus \{\pi\}$, either $\sum_{i \in T''_\rho \cap S_{\rho\sigma}}|x_i| < |e_{\rho\sigma}|$ and $\sigma \in E'_\rho$ or $T''_{\rho} \cap S_{\rho\sigma}=\emptyset$.
Using these facts, it follows from (\ref{BigBound}) and (\ref{qBoundEq}) that $\ell_{\rho}(j)=-|z_{\pi\rho}|-\frac{1}{2}(q-||{\bf z}_\rho||_1)>-(s+\epsilon)$.

\emph{Proof of (iii).}
Let $\rho$ index a row of $P$ that separates $T^+\setminus \{j\}$ from $T^-\cup \{j\}$ and let $S_{\rho\pi}$ be the partition class induced by row $\rho$ of $P$ that contains $j$.
If $z_{\rho\pi} < 0$, then $\sum_{i \in T''_\rho \cap S_{\rho\pi}}|x_i| \leq x_j$.
Furthermore, for each symbol $\sigma \in \{1,\ldots,k_\rho\} \setminus \{\pi\}$, either $\sum_{i \in T''_\rho \cap S_{\rho\sigma}}|x_i| \leq |e_{\rho\sigma}|$ and $\sigma \in E'_\rho$ or $T''_{\rho} \cap S_{\rho\sigma}=\emptyset$.
Then, it follows from (\ref{MisplacedEq}) and (\ref{qBoundEq}) that $u_\rho(j) = \frac{1}{2}(q-||{\bf z}_\rho||_1)-(s+\epsilon) < x_j$.
If $z_{\rho\pi} \geq 0$, then $\sum_{i \in T' \cap S_{\rho\pi}}|x_i| \leq x_j$.
Furthermore, for each symbol $\sigma \in \{1,\ldots,k_\rho\} \setminus \{\pi\}$, either $\sum_{i \in T''_\rho \cap S_{\rho\sigma}}|x_i| \leq |e_{\rho\sigma}|$ and $\sigma \in E'_\rho$ or $T''_{\rho} \cap S_{\rho\sigma}=\emptyset$.
Then, it follows from (\ref{BigBound}) and (\ref{qBoundEq}) that $u_\rho(j) = |z_{\pi\rho}|+\frac{1}{2}(q-||{\bf z}_\rho||_1)-(s+\epsilon)<x_j$.
\end {IEEEproof}


\section{Conclusion} \label{sec:conclusion}

Hierarchical construction of measurement matrices by column replacement permits the explicit construction of large measurement matrices from small ones.
The use of heterogeneous hash families supports the use of a library of smaller ingredient matrices, while the use of strengthening hash families allows the ingredient matrices to be designed for lower sparsity than the larger measurement matrix produced.
Perhaps surprisingly, the ingredient measurement matrices need not all employ the same recovery algorithm; rather recovery for the large measurement matrix can use arbitrary routines for recovery that are provided with the ingredient matrices.
In this way, computationally intensive recovery methods can be used for the ingredient matrices, which permits the selection of smaller matrices in general, while still enabling recovery for the large measurement matrix.
Nevertheless, recovery using the large measurement matrix can be computationally prohibitive without further restrictions.
Therefore it is shown that using a standard construction of linear hash families over the finite field, recovery for the large measurement matrix can be effected in sublinear time.
Indeed sublinear recovery time can be obtained even when computationally intensive methods are used for each ingredient matrix.
A practical implementation of these recovery methods requires that the methods deal effectively with noise in the signal.
Suitable restrictions on the hash family and on each ingredient matrix used in column replacement are shown to be sufficient to permit recovery even in the presence of such noise.

Measurement matrices that result from one column replacement have been studied here.
Because recovery does not depend on the method by which recovery is done for the ingredient matrices, it is possible that the ingredient matrices themselves are constructed by column replacement from even smaller ingredient matrices.
The merits and demerits of repeated column replacement deserve further study.


\section*{Acknowledgements}

The work of D. Horsley and C.~J. Colbourn is supported in part by the Australian Research Council through grant DP120103067.


\bibliographystyle{abbrv}
\bibliography{IEEEabrv,references}

\begin{thebibliography}{10}

\bibitem{Alon86}
N.~Alon.
\newblock Explicit construction of exponential sized families of
  {$k$}-independent sets.
\newblock {\em Discrete Mathematics}, 58:191--193, 1986.

\bibitem{Baraniuk}
R.~Baraniuk.
\newblock Compressive sensing.
\newblock {\em IEEE Signal Processing Magazine}, 24:227--234, 2007.

\bibitem{Berinde}
R.~Berinde, A.~C. Gilbert, P.~Indyk, H.~Karloff, and M.~J. Strauss.
\newblock Combining geometry and combinatorics: A unified approach to sparse
  signal recovery.
\newblock In {\em Proceedings of the 46th Annual Allerton Conference on
  Communication, Control, and Computing}, pages 798--805, 2008.

\bibitem{BESZ}
S.~R. Blackburn, T.~Etzion, D.~R. Stinson, and G.~M. Zaverucha.
\newblock A bound on the size of separating hash families.
\newblock {\em Journal of Combinatorial Theory, Series A}, 115((7):1246--1256,
  2008.

\bibitem{BlackburnWild}
S.~R. Blackburn and P.~R. Wild.
\newblock Optimal linear perfect hash families.
\newblock {\em Journal of Combinatorial Theory, Series A}, 83:233--250, 1998.

\bibitem{candes06-1}
E.~J. Cand{\`e}s.
\newblock Compressive sampling.
\newblock In {\em International Congress of Mathematicians}, volume~3, pages
  1433--1452, 2006.

\bibitem{candes08}
E.~J. Cand{\`e}s.
\newblock The restricted isometry property and its implications for compressed
  sensing.
\newblock {\em Compte Rendus de l'Academie des Sciences, Series I},
  346:589--592, 2008.

\bibitem{candes06-3}
E.~J. Cand{\`e}s, J.~Romberg, and T.~Tao.
\newblock Robust uncertainty principles: Exact signal reconstruction from
  highly incomplete frequency information.
\newblock {\em IEEE Transactions on Information Theory}, 52:489--509, 2006.

\bibitem{candes05}
E.~J. Cand{\`e}s and T.~Tao.
\newblock Decoding by linear programming.
\newblock {\em IEEE Transactions on Information Theory}, 51:4203--4215, 2005.

\bibitem{candes06-2}
E.~J. Cand{\`e}s and T.~Tao.
\newblock Near optimal signal recovery from random projections: Universal
  encoding strategies.
\newblock {\em IEEE Transactions on Information Theory}, 52:5406--5425, 2006.

\bibitem{ChenDS}
S.~S. Chen, D.~L. Donoho, and M.~A. Saunders.
\newblock Atomic decomposition by basis pursuit.
\newblock {\em SIAM Journal on Scientific Computing}, 20(1):33--61, 1998.

\bibitem{Cohen}
A.~Cohen, W.~Dahmen, and R.~A. DeVore.
\newblock Compressed sensing and best $k$-term approximation.
\newblock {\em Journal of the American Mathematical Society}, 22:211--231,
  2009.

\bibitem{CHM-TCom}
C.~J. Colbourn, D.~Horsley, and C.~Mc{L}ean.
\newblock Compressive sensing matrices and hash families.
\newblock {\em IEEE Transactions on Communications}, 59(7):1840--1845, 2011.

\bibitem{ColbournHorsleySyrotiuk11}
C.~J. Colbourn, D.~Horsley, and V.~R. Syrotiuk.
\newblock Strengthening hash families and compressive sensing.
\newblock {\em Journal of Discrete Algorithms}, 16:170--186, 2012.

\bibitem{CLlinhf}
C.~J. Colbourn and A.~C.~H. Ling.
\newblock Linear hash families and forbidden configurations.
\newblock {\em Designs, Codes and Cryptography}, 59:25--55, 2009.

\bibitem{Graham}
G.~Cormode and S.~Muthukrishnan.
\newblock Combinatorial algorithms for compressed sensing.
\newblock In {\em Lecture Notes in Computer Science}, volume 4056, pages
  280--294, 2006.

\bibitem{devore07-1}
R.~A. DeVore.
\newblock Deterministic constructions of compressed sensing matrices.
\newblock {\em Journal of Complexity}, 23:918--925, 2007.

\bibitem{Donoho01}
D.~L. Donoho and X.~Huo.
\newblock Uncertainty principles and ideal atomic decomposition.
\newblock {\em IEEE Transactions on Information Theory}, 47:2845--2862, 2001.

\bibitem{EladBruckstein}
M.~Elad and A.~M. Bruckstein.
\newblock A generalized uncertainty principle and sparse representation in
  pairs of bases.
\newblock {\em IEEE Transactions on Information Theory}, 48:2558--2567, 2002.

\bibitem{Fuchs2004}
J.~J. Fuchs.
\newblock On sparse representations in arbitrary redundant bases.
\newblock {\em IEEE Transactions on Information Theory}, 50:1341--1344, 2004.

\bibitem{Fuchs2005}
J.~J. Fuchs.
\newblock Recovery of exact sparse representations in the presence of bounded
  noise.
\newblock {\em IEEE Transactions on Information Theory}, 51:3601--3608, 2005.

\bibitem{Gilbert}
A.~C. Gilbert, M.~A. Iwen, and M.~J. Strauss.
\newblock Group testing and sparse signal recovery.
\newblock In {\em Proceedings of the 42nd Asilomar Conference on Signals,
  Systems}, pages 1059--1063, 2008.

\bibitem{Gilbert01}
A.~C. Gilbert, M.~J. Strauss, J.~Tropp, and R.~Vershynin.
\newblock One sketch for all: Fast algorithms for compressed sensing.
\newblock In {\em Proceedings of the ACM Symposium on Theory of Computing},
  pages 237--246, 2007.

\bibitem{GribonvalMorten2003}
R.~Gribonval and M.~Nielsen.
\newblock Sparse representations in unions of bases.
\newblock {\em IEEE Transactions on Information Theory}, 49:3320--3325, 2003.

\bibitem{Iwenpreprint}
M.~A. Iwen.
\newblock Combinatorial sublinear-time {F}ourier algorithms.
\newblock {\em Foundations of Computational Mathematics}, 10:303--338, 2010.

\bibitem{JafarpourEtAl2009}
S.~Jafarpour, W.~Xu, B.~Hassibi, and R.~Calderbank.
\newblock Efficient and robust compressed sensing using optimized expander
  graphs.
\newblock {\em IEEE Transactions on Information Theory}, 55:4299--4308, 2009.

\bibitem{Natarajan}
B.~K. Natarajan.
\newblock Sparse approximate solutions to linear systems.
\newblock {\em SIAM Journal on Computing}, 24:227--234, 1995.

\bibitem{STW}
D.~R. Stinson, {Tran Van Trung}, and R.~Wei.
\newblock Secure frameproof codes, key distribution patterns, group testing
  algorithms and related structures.
\newblock {\em Journal of Statistical Planning and Inference}, 86:595--617,
  2000.

\bibitem{StinsonWC08}
D.~R. Stinson, R.~Wei, and K.~Chen.
\newblock On generalized separating hash families.
\newblock {\em Journal of Combinatorial Theory, Series A}, 115:105--120, 2008.

\bibitem{Stojnic}
M.~Stojnic, W.~Xu, and B.~Hassibi.
\newblock Compressed sensing-probabilistic analysis of a null-space
  characterization.
\newblock In {\em Proceedings of the International Conference on Acoustics,
  Speech, and Signal Processing}, pages 3377--3380, 2008.

\bibitem{Tropp2005}
J.~A. Tropp.
\newblock Recovery of short, complex linear combinations via {$l_1$}
  minimization.
\newblock {\em IEEE Transactions on Information Theory}, 51:1568--1570, 2005.

\bibitem{XuHassibi2007}
W.~Xu and B.~Hassibi.
\newblock Efficient compressive sensing with deterministic guarantees using
  expander graphs.
\newblock In {\em Proceedings of IEEE Information Theory Workshop}, 2007.

\bibitem{Zhang08}
Y.~Zhang.
\newblock On theory of compressive sensing via $\ell_1$-minimization: Simple
  derivations and extensions.
\newblock Technical Report Technical Report CAAM TR08-11, Rice University,
  2008.

\end{thebibliography}

\end{document}